\documentclass[11pt,titlepage,a4paper]{article}
\usepackage{bozhomac}	
\usepackage{bozhlogo}	
\usepackage{cite}

\title{\bfseries	\vspace*{-1.678902345in}
\vspace*{-7ex}
{
\begin{flushright}
     \textbf{\large LANL arXiv server, E-print No. physics/0109033}\\[5ex]
\end{flushright}
}
{\huge On fibre bundle formulation of classical and statistical mechanics}
}

\vspace{1.7ex}

\author{
Bozhidar Z. Iliev
\thanks{Department Mathematical Modeling,
Institute for Nuclear Research and \mbox{Nuclear} Energy,
Bulgarian Academy of Sciences,
Boul. Tzarigradsko chauss\'ee~72, 1784 Sofia, Bulgaria}
\thanks{E-mail address: bozho@inrne.bas.bg}
\thanks{URL: http://theo.inrne.bas.bg/$\sim$bozho/}
}

\date{		
 \vspace{2.27ex}\ShortTitle{On bundle classical and  statistical mechanics}
								\\[0.27ex]
 \vspace{3.27ex}
	\begin{tabular}{r@{$\colon\to~$}l}
 \vspace{0.09ex} Basic ideas 	& October 1997		\\[0.09ex]
 \vspace{0.09ex} Began	 	& November 6, 1997	\\[0.09ex]
 \vspace{0.09ex} Ended	 	& December 20, 1997	\\[0.09ex]
 \vspace{0.09ex} Revised	& August 1999	\\[0.09ex]
 \vspace{0.09ex } Last update	& September 16, 2001	\\[0.09ex]
 \vspace{0.27ex} Produced 	&\fbox{\today}	\\[0.27ex]
	\end{tabular} \\[1.27ex]
	\begin{tabular}{r@{$\colon~$}l}
 \vspace{0.27ex} LANL xxx archive server E-print No. & physics/0109033
							\\[0.27ex]
	\end{tabular} \\[-0.27ex]
 \vspace{4.27ex}{\Huge\BOZHO}		\\[4.27ex]
 \vspace{0.27ex}\Subject{Classical mechanics, Differential geometry}\\[2.27ex]
	\begin{tabular}{r@{\hspace{0.512em}}|@{\hspace{0.512em}}l}
 \vspace{0.27ex} 
&
 \vspace{0.27ex} 
	\end{tabular} \\[1.27ex]
 \vspace{0.27ex}	\\[0.27ex]
}	

\newcommand{\phasespace}{\mathcal{F}}		
\newcommand{\bundle}{(\total,\proj,\base)}	
	\newcommand {\total} {\mathit{R}}	
	\newcommand {\proj}  {\pi_\total}	
	\newcommand {\base}  {\mathit{M}}	

\newcommand{\Ham}   {\mathit{h}}		
\newcommand{\Liouv} {\mathcal{L}}	
\newcommand{\distr} {\mathcal{P}}	
\newcommand{\bdistr}{\mathit{P}}	

\newcommand{\dyn}[1]{\mathsf{#1}}		
	\newcommand{\ope}[1]{\mathcal{#1}}		 
	\newcommand{\mor}[1]{\mathit{#1}}		 

\newcommand{\Poisson}[2]{\left[ #1,#2 \right]_\mathrm{P}}
\newcommand{\poisson}[1]{\left[ #1 \right]_\mathrm{P}}	 

\begin{document}		

\renewcommand{\thefootnote}{\fnsymbol{footnote}} 
\maketitle				
\renewcommand{\thefootnote}{\arabic{footnote}}   

\tableofcontents		


\pagestyle{myheadings}
\markright{\itshape\bfseries Bozhidar Z. Iliev:
\upshape\sffamily\bfseries On bundle classical and  statistical mechanics}

\begin{abstract}

	Some elements of classical mechanics and classical statistical
mechanics are formulated in terms of fibre bundles. In the bundle approach
the dynamical and distribution functions are replaced by liftings of paths in
a suitably chosen bundle. Their time evolution is described by
appropriate linear transports along paths in it or, equivalently, by
corresponding invariant bundle equations of motion. In particular, the
bundle version of the Liouville equation is derived.

\end{abstract}

\section {Introduction}
\label{Sect1}
\setcounter{equation} {0}

	In the series of papers~\cite{bp-BQM-introduction+transport,
bp-BQM-equations+observables, bp-BQM-pictures+integrals,
bp-BQM-mixed_states+curvature, bp-BQM-interpretation+discussion},
we have reformulated nonrelativistic quantum mechanics in terms of fibre
bundles. In the present work, we want to try to apply some ideas and methods
from these papers to classical mechanics and classical statistical mechanics.
However, as a whole this is scarcely possible because these theories are more
or less primary related to the theory of space (space-time) which is taken as
a base of the corresponding bundle(s) in the bundle approach and,
consequently, it has to be determined by other theory. By this reason, the
fibre bundle formalism is only partially applicable to some elements of
classical mechanics and classical statistical mechanics.

	A different geometrical approach to the statistical mechanics, based
on the projective geometry, can be found in~\cite{Brody&Hughston}.

	The organization of this paper is the following. In
Sect.~\ref{Sect2} are recalled some facts of classical Hamiltonian mechanics
and fix our notation. In Sect.~\ref{Sect3}, we give a fibre bundle
description of (explicitly time-independent) dynamical functions,
representing the observables in classical mechanics. In this approach they
are represented by liftings of paths in a suitably chosen bundle. We show
that their time evolution is governed by a kind of linear (possibly parallel)
transport along paths in this bundle or, equivalently, via the corresponding
bundle equation of motion derived here.  Sect.~\ref{Sect4} is devoted to the
bundle (analogue of the) Liouville equation, the equation on which classical
statistical mechanics rests. In the bundle description, we replace the
distribution function by a lifting of paths in the same bundle appearing in
Sect.~\ref{Sect3}. In it we derive the bundle version of the Liouville
equation which  turns to be the equation for (linear) transportation of this
lifting with respect to a suitable linear transport along paths. The paper
closes with some remarks in Sect.~\ref{Concl}.

\section[Hamilton description of classical mechanics   (review)]
	{Hamilton description of classical mechanics \\(review)}
\label{Sect2}
\setcounter{equation} {0}

	In classical mechanics~\cite{Goldstein-CM} the state of a dynamical
system is accepted to be describe via its generalized coordinates
$q=(q_1,\ldots,q_N)\in\mathbb{R}^N$ and momenta
$p=(p_1,\ldots,p_N)\in\mathbb{R}^N$ with $N\in\mathbb{N}$ being the number of
system's degree of freedom.
The quantities characterizing a dynamical system, the so called
dynamical functions or variables, are described by $C^1$ functions in
 $\mathbb{R}^\phasespace = \{ \dyn{f}\colon\phasespace\to\mathbb{R} \}$ with
$\phasespace$ being the system's phase space. The Poisson bracket of
$\dyn{f},\dyn{g}\in\mathbb{R}^\phasespace$ is~\cite[\S~8.4]{Goldstein-CM}
	\begin{equation}	\label{2.3}
\Poisson{\dyn{f}}{\dyn{g}} :=
\sum_{i=1}^{N}
\left(
\frac{\partial\dyn{f}}{\partial q^i} \frac{\partial\dyn{g}}{\partial p_i}
-
\frac{\partial\dyn{f}}{\partial p_i} \frac{\partial\dyn{g}}{\partial q^i}
\right)
	\end{equation}
which is an element of $\mathbb{R}^\phasespace$. The subset of
$\mathbb{R}^\phasespace$ consisting of $C^1$ functions and endowed with the
operations addition, multiplication (with real numbers) and forming of
Poisson brackets is called \emph{dynamical algebra} and will be denoted
by $\mathcal{D}$~\cite[Section~1.2]{Balescu}. The set $\mathcal{D}$ is closed
with respect to the mentioned operations and is a special kind of Lie
algebra, the Poisson bracket playing the r\^ole of Lie bracket.

	If $\Ham(q,p;t)$ is the system's Hamiltonian,
the system evolves in time $t\in\mathbb{R}$ according to the (canonical)
Hamilton equations~\cite[chapter~7, \S~8.5]{Goldstein-CM}
	\begin{equation}	\label{2.4}
\dot q^{i} =   \frac{\partial\Ham(q,p;t)}{\partial p_i}
 = \Poisson{q^i}{\Ham}, \quad
\dot p_{i} = - \frac{\partial\Ham(q,p;t)}{\partial q^i}
 = \Poisson{p_i}{\Ham},
	\end{equation}
where $i=1,\ldots,N$ and the dot means full derivative with respect to time,
e.g.
$\dot q^{i}:=\ordinary q^i/\ordinary t$.
The system's state is completely known for every
instant $t$ if for some $t_0\in\mathbb{R}$ are fixed the initial values
 $(q,p)|_{t=t_0}=(q_0,p_0)\in\phasespace$ with
 $\phasespace=\mathbb{R}^{2N}$ being system's phase space.

	If $g$ is depending on time dynamical function,
$\dyn{g}\in\mathbb{R}^{\phasespace\times\mathbb{R}}$,
then its full time derivative is~\cite[equation~(8.58)]{Goldstein-CM}
	\begin{equation}	\label{2.5}
\frac{\ordinary \dyn{g}}{\ordinary t} =: \dot{\dyn{g}} =
	\Poisson{\dyn{g}}{\Ham} +\frac{\partial\dyn{g}}{\partial t}.
	\end{equation}

	To any dynamical function $\dyn{f}\in\mathcal{D}$ there corresponds
operator
$\poisson{\dyn{f}}\colon\dyn{g}\mapsto\Poisson{\dyn{g}}{\dyn{f}}$
$\dyn{g}\in\mathcal{D}$,
\ie
	\begin{equation}	\label{2.6}
\poisson{\dyn{f}} := \Poisson{\cdot}{\dyn{f}}\colon \mathcal{D}\to\mathcal{D}.
	\end{equation}

	Putting
$\xi:=(q,p)=(q^1,\ldots,q^N,p_{1},\ldots,p_{N})\in\phasespace$
and defining the map
 $\overline{\Ham} \colon \phasespace\to\phasespace$ by
\(
\overline{\Ham}\colon (q,p)\mapsto
(\poisson{\Ham}q^1,\ldots,\poisson{\Ham}q^N,
\poisson{\Ham}p_1,\ldots,\poisson{\Ham}p_N) ,
\)
which map can be called \emph{Hamiltonian operator},
we see that~\eref{2.4} is equivalent to
	\begin{equation}	\label{2.7}
\frac{\ordinary \xi}{\ordinary  t} = \overline{\Ham}(\xi) .
	\end{equation}

\section[Bundle description of dynamical functions\\ in classical mechanics]
	{Bundle description of dynamical functions\\ in classical mechanics}
\label{Sect3}
\setcounter{equation} {0}

	At first sight, it seems the solution of~\eref{2.7} might be
written as $\xi(t)=\ope{U}(t,t_0)\xi(t_0)$ with $\ope{U}(t,t_0)$ being the
Green's function for this equation. However, this is wrong as generally
$\Ham$ depends on $\xi$, $\Ham=\Ham(\xi;t)$, so $\ope{U}$ itself must depend
on $\xi$. Consequently, we cannot apply to the Hamiltonian
equation~\eref{2.7} the developed in~\cite{bp-BQM-introduction+transport}
method for fibre bundle interpretation and reformulation of Schr\"odinger
equation. The basic reason for this is that the Hamilton equation is primary
related to the (phase) space while the Schr\"odinger one is closely related
to the `space of observables'. This suggests the idea of bundle description
of dynamical functions which are the classical analogue of quantum
observables. Below we briefly realize it for time-independent dynamical
functions.

	Let $\dyn{g}\in\mathcal{D}$ and $\partial\dyn{g}/\partial t =0$.
By~\eref{2.5} and~\eref{2.6}, we have
	\begin{equation}	\label{3.1}
\ordinary \dyn{g}/\ordinary  t = \poisson{h}\dyn{g} .
	\end{equation}
Writing for brevity $\dyn{g}(t)$ instead of
$\dyn{g}(\xi(t);t)=\dyn{g}(\xi(t);t_0)$, we can put
	\begin{equation}	\label{3.2}
\dyn{g}(t) = \ope{V}(t,t_0)\dyn{g}(t_0),
	\end{equation}
where $t_0$ is a fixed instant of time and the \emph{dynamical operator}
$\ope{V}$, the Green function of~\eref{3.1}, is defined via the
initial-value problem
	\begin{equation}	\label{3.3}
\frac{\partial\ope{V}(t,t_0)}{\partial t} = \poisson{h}\ope{V}(t,t_0), \qquad
\ope{V}(t_0,t_0) = \boldsymbol{1}.
	\end{equation}
(Here $\boldsymbol{1}$ is the corresponding unit operator.)

	The explicit form of $\ope{V}(t,t_0)$ is
	\begin{equation}	\label{3.4}
\ope{V}(t,t_0) =
\left.
\left( \Texp\int\limits_{t_0}^{t}\poisson{h(\xi;\tau)} \od\tau\, \right)
\right|_{\xi=\xi(t_0)}
	\end{equation}
where $\Texp\int_{t_0}^{t}$\ldots denotes the chronological (called
also T-ordered, P-ordered, or path-ordered) exponent. One can easily check
the linearity of $\ope{V}(t,t_0)$ and the equalities
	\begin{align}
\ope{V}(t_3,t_1) &= \ope{V}(t_3,t_2)\ope{V}(t_2,t_1),  	\label{3.5}	\\
\ope{V}(t_1,t_1) &= \boldsymbol{1},		       	\label{3.6}	\\
\ope{V}^{-1}(t_1,t_2) &= \ope{V}(t_2,t_1),		\label{3.7}
	\end{align}
the last of which is a consequence of the preceding two. Here $t_1$,
$t_2$ and $t_3$ are any three moments of time.

	Let $\base$ and $\mathbb{T}$ be the classical Newtonian respectively
3-dimensional space and 1-dimensional time of classical mechanics.%
\footnote{%
$\base$ and $\mathbb{T}$ are isomorphic to $\mathbb{R}^3$ and $\mathbb{R}^1$
respectively. This is insignificant for the following.%
}
Let $\gamma\colon J\to\base,\ J\subseteq\mathbb{T}$, be the trajectory of
some (point-like) observer (if the observer exists for all $t\in\mathbb{T}$,
then $J=\mathbb{T}$.)

	Now define a bundle $\bundle$ with a total space $\total$, base
$\base$, projection $\proj\colon\total\to\base$, and isomorphic fibres
$\total_x:=\proj^{-1}(x)=d_{x}^{-1}(\mathbb{R})$ where $\mathbb{R}$ is
regarded as a standard fibre of $\bundle$ and
$d_x\colon\total_x\to\mathbb{R}$ are (arbitrarily) fixed isomorphisms.

	To every function
$\dyn{g}\colon\phasespace\times\mathbb{T}\to\mathbb{R}$,
we assign a lifting of paths%
\footnote{%
Equivalently, the mapping $g$ can be regarded as a (multiple\ndash valued)
section along paths; see~\cite[sect.~3 \&4]{bp-BQM-introduction+transport}.%
}
$\mor{g}$ such that
	\begin{equation}	\label{3.8}
\mor{g}\colon\gamma\mapsto \mor{g}_\gamma
	\colon t \mapsto\mor{g}_\gamma(\xi;t) :=
d_{\gamma(t)}^{-1}(\dyn{g}(\xi;t)) \in\total_{\gamma(t)}.
	\end{equation}
In this way the dynamical algebra $\mathcal{D}$ becomes isomorphic to a
subalgebra of the algebra of liftings of paths (or sections along paths) of
$\bundle$.

	For explicitly time-independent dynamical
functions, substituting~\eref{3.8} into~\eref{3.2}, we get
	\begin{gather}
\mor{g}_\gamma(t) = \mor{V}_\gamma(t,t_0) \mor{g}_\gamma(t_0),
							\label{3.9}	\\
\intertext{where, for brevity, we write
	$g_\gamma(t):=g_\gamma(\xi(t);t)=g_\gamma(\xi(t);t_0)$ and
	  }
\mor{V}_\gamma(t,t_0) :=
d_{\gamma(t)}^{-1}\circ\ope{V}(t,t_0)\circ d_{\gamma(t_0)}
\ : \ \total_{\gamma(t_0)}\to\total_{\gamma(t)} .
							\label{3.10}
	\end{gather}
The map $\mor{V}_\gamma(t,t_0)$ is linear and, due to~\eref{3.5}
and~\eref{3.6}, satisfies the equations
	\begin{align}
\mor{V}(t_3,t_1) & = \mor{V}(t_3,t_2)\mor{V}(t_2,t_1),  \label{3.11}	\\
\mor{V}(t_1,t_1) & = \boldsymbol{1}.			\label{3.12}
	\end{align}
The last three equations show that
\(
\mor{V} \colon \gamma \mapsto \mor{V}_\gamma \colon(t,t_0) \mapsto
\mor{V}_\gamma(t,t_0)
\)
is a linear transport along paths in $\bundle$
(cf.~\cite{bp-BQM-introduction+transport} or~\cite{bp-LTP-general}).
We call it the \emph{dynamical transport}.

	By~\cite[proposition~5.3]{bp-LTP-appl}
or~\cite[eq.~(3.40)]{bp-BQM-introduction+transport}, equation~\eref{3.9} is
equivalent to
	\begin{equation}	\label{3.13}
{^\mor{V}\negthickspace}\mor{D}(\mor{g}) = 0
	\end{equation}
Here ${^\mor{V}\negthickspace}\mor{D}$ is the derivation along paths
corresponding to $\mor{V}$, viz. (see~\cite[definition~4.1]{bp-LTP-general},
\cite{bp-normalF-LTP},
or~\cite[definition~3.4]{bp-BQM-introduction+transport})
\[
{^\mor{V}\negthickspace}\mor{D} \colon
 	\PLift^1\bundle \to \PLift^0\bundle
\]
where $\PLift^k\bundle$, $k=0,1,\ldots$ is the set of $C^k$ liftings of paths
from $\base$ to $\total$, and its action on a lifting
$\lambda\in\PLift^1\bundle$ with $\lambda\colon\gamma\mapsto\lambda_\gamma$
is given via
	\begin{equation}	\label{3.14}
	{^\mor{V}\negthickspace}\mor{D}_{t}^{\gamma} (\lambda) :=
\lim_{\varepsilon\to0}
\left\{
\frac{1}{\varepsilon}
\left[
\mor{V}_\gamma(t,t+\varepsilon)\lambda_\gamma(t+\varepsilon)
	- \lambda_\gamma(t)
\right]
\right\}
	\end{equation}
with
\(
{^\mor{V}\negthickspace}\mor{D}^\gamma_t (\lambda)
 := (({^\mor{V}\negthickspace}\mor{D} \lambda) (\gamma)) (t)
 = ( {^\mor{V}\negthickspace}\mor{D} \lambda)_\gamma (t) .
\)

	The equivalence of~\eref{3.13} and the conventional equation of
motion~\eref{3.1} can easily be verified. Therefore~\eref{3.13} represents
the \emph{bundle equation of motion} for dynamical functions.

	To conclude, we emphasize on the fact that the application of the
bundle approach, developed
in~\cite{bp-BQM-introduction+transport,bp-BQM-equations+observables},
to classical mechanics results only in bundle  description of dynamical
functions.

\section {Bundle description of the Liouville equation}
\label{Sect4}
\setcounter{equation} {0}

	In classical statistical mechanics~\cite{Balescu} the
evolution of a system is described via a distribution (function (on the phase
space)) $\distr\colon\phasespace\times\mathbb{T}\to\mathbb{R}$ satisfying the
conditions
\(
\int\limits_{\phasespace}^{}\od\xi\, \distr(\xi;t) = 1\text{ and }
\distr(\xi;t)\ge 0,\ \xi\in\phasespace,\  t\in\mathbb{T},
\)
and whose time evolution is governed by the Liouville equation
	\begin{equation}	\label{4.2}
\frac{\partial\distr}{\partial t} = \Liouv\distr
	\end{equation}
with $\Liouv=\Liouv(\xi;t)$ being the Liouville operator (the Liouvillian) of
the investigated system~\cite[\S~2.2]{Balescu}. If the system is Hamiltonian,
\ie  if it can be described via a Hamiltonian $\Ham$, its Liouvillian is
$\Liouv=-\poisson{\Ham}$.

	Since equations~\eref{3.1} and~\eref{4.2} are similar,  we can apply
the already developed ideas and methods to the bundle reformulation of the
basic equation of classical statistical mechanics.

	We can write the solution of~\eref{4.2} as
	\begin{equation}	\label{4.4}
\distr(\xi;t)=\ope{W}(\xi;t,t_0)\distr(\xi;t_0)
	\end{equation}
where the \emph{distribution operator} $\ope{W}$ is defined by the
initial-value problem
	\begin{equation}	\label{4.5}
\frac{\partial\ope{W}(\xi;t,t_0)}{\partial t} =
\Liouv(\xi;t)\ope{W}(\xi;t,t_0),\quad \ope{W}(\xi;t_0,t_0) = \boldsymbol{1},
	\end{equation}
\ie
\(
\ope{W}(\xi;t,t_0)
 =\Texp\int\limits_{t_0}^{t}\Liouv(\xi;\tau)\,\od\tau.
\)

	Since $\ope{W}$ satisfies~\eref{3.5} and~\eref{3.6} with $\ope{W}$
instead of $\ope{V}$, a fact that can easily be checked, the maps
	\begin{equation}	\label{4.6}
\mor{W}(\xi;t,t_0) :=
d_{\gamma(t)}^{-1}\circ\ope{W}(\xi;t,t_0)\circ d_{\gamma(t_0)}
	\colon \total_{\gamma(t_0)}\to\total_{\gamma(t)}
	\end{equation}
satisfies~\eref{3.11} and~\eref{3.12}. Therefore these maps define a
transport $\mor{W}$ along paths in $\bundle$. It can be called the
\emph{distribution transport}.

	Now to any distribution
$\distr\colon\phasespace\times\mathbb{T}\to\mathbb{R}$, we assign a
(distribution) lifting $\bdistr$ of paths in the fibre bundle $\bundle$,
introduced in Sect.~\ref{Sect3}, such that
	\begin{equation}	\label{4.7}
\bdistr\colon\gamma\mapsto \bdistr_\gamma \colon
	t\mapsto\bdistr_\gamma(\xi;t)
:=	d_{\gamma(t)}^{-1}(\distr(\xi;t)) \in\total_{\gamma(t)}.
	\end{equation}
The so-defined lifting
$\bdistr\colon\gamma\to\bdistr_\gamma$ of paths in
$\bundle$ is linearly transported along arbitrary observer's trajectory
$\gamma$ by means of $\mor{W}$. In fact, combining~\eref{4.4}
and~\eref{4.7}, using~\eref{4.7} for $t=t_0$ and~\eref{4.6}, we get
	\begin{equation}	\label{4.8}
\bdistr_\gamma(\xi;t) = \mor{W}_\gamma(\xi;t,t_0) \bdistr_\gamma(\xi;t_0)
	\end{equation}
which proves our assertion. We want to emphasize on the equivalence
of~\eref{4.8} and the Liouville equation~\eref{4.2}, a fact following from
the derivation of~\eref{4.8} and the definitions of the quantities appearing
in it. This result, combined with~\cite[proposition~5.3]{bp-LTP-appl} shows
the equivalence of~\eref{4.2} with the invariant equation
	\begin{equation}	\label{4.9}
{^\mor{W}\negthickspace}D(\bdistr) = 0
	\end{equation}
where
${^\mor{W}\negthickspace}D$ is the derivation along $\gamma$
corresponding to $\mor{W}$ (see~\eref{3.14}). The last equation can naturally
be called the \emph{bundle Liouville equation}.

\section {Conclusion}
\label{Concl}
\setcounter {equation} {0}

	In this paper we tried to apply the methods developed
in~\cite{bp-BQM-introduction+transport,
bp-BQM-equations+observables, bp-BQM-pictures+integrals,
bp-BQM-mixed_states+curvature, bp-BQM-interpretation+discussion}
for quantum mechanics to classical mechanics and classical statistical
mechanics. Regardless that these methods are fruitful in quantum
mechanics, they do not work with the same effectiveness in classical
mechanics and statistics. The main reason for this is that these mechanics are
more or less theories of space (and time), \ie they directly depend on the
accepted space (and time) model. So, since the fibre bundle formalism, we are
attempting to transfer from quantum mechanics and statistical to classical
ones, is suitable for describing quantities directly
insensitive to the space(-time) model, we can realize the ideas
of~\cite{bp-BQM-introduction+transport,
bp-BQM-equations+observables, bp-BQM-pictures+integrals,
bp-BQM-mixed_states+curvature, bp-BQM-interpretation+discussion}
in the classical region only partially.

	In this work we represented dynamical and distribution functions as
liftings of paths of a suitably chosen fibre bundle over space. These
liftings, as it was demonstrated, appear to be linearly transported along any
observer's trajectory with respect to corresponding (possibly parallel)
transports along paths in the bundle mentioned. As a consequence of this
fact, the equations of motion for distributions and time-independent
dynamical functions have one and the same mathematical form: the derivations,
generated by the corresponding transports, of these liftings vanish along
observer's trajectory.

	Thus, we have seen that (some) quantities arising over space admit
natural bundle formulation which is equivalent to the conventional one. We
demonstrated this for time-independent dynamical functions in classical
Hamiltonian mechanics and distribution functions in classical statistical
mechanics.  Other classical quantities also admit bundle description.

	The fibre bundle formalism is extremely suitable for describing all
sorts of fields over space(\ndash time). Therefore it seems naturally
applicable to quantum physics. In particular, this is true for
nonrelativistic and relativistic quantum mechanics (and statistics) whose
full self-consistent bundle (re)formulation we have developed in the series
of papers~\cite{bp-BQM-introduction+transport, bp-BQM-equations+observables,
bp-BQM-pictures+integrals, bp-BQM-mixed_states+curvature,
bp-BQM-interpretation+discussion,bp-BRQM-time-dependent,bp-BRQM-covariant}.

\addcontentsline{toc}{section}{References}
\bibliography{bozhopub,bozhoref}
\bibliographystyle{unsrt}
\addcontentsline{toc}{subsubsection}{This article ends at page}

\end{document}